\documentclass[]{spie}  %>>> use for US letter paper

\usepackage{amsmath,amsfonts,amssymb, bbold}
\usepackage{graphicx}
\usepackage[colorlinks=true, allcolors=blue]{hyperref}
\usepackage{cite}
\usepackage{mathrsfs}
\usepackage{subcaption}
\usepackage{changes}

\title{First laboratory demonstration of iEFC for high-contrast imaging at small angular separation on ELT scales}
\author[a] {C. Sallard}
\author[a] {P. Martinez}
\author[a] {A. Spang}
\author[b] {K. Milani}
\author[a] {L. Abe}
\author[a] {M. Beaulieu}
\author[a] {C. Gouvret}
\author[a] {A. Marcotto}
\author[a] {J. Dejonghe}
\author[a] {O. Preis}
\affil[a]{Universit\'e C\^ote d'Azur, Observatoire de la C\^ote d'Azur, CNRS, Laboratoire Lagrange, France}
\affil[b]{Stewart Observatory, Univ. of Arizona, United States}

\authorinfo{Send correspondence to C. Sallard, E-mail: Charles.Sallard@oca.eu}

\begin{document} 
\maketitle
\begin{abstract}
% C'est l'abstract qui a été soumis à SPIE
Implicit electric field conjugation (iEFC) is an efficient, model-free wavefront control (WFC) technique for dark-hole (DH) creation.
Its model-free nature, relying on calibration with real data rather than a synthetic model of the instrument, makes it uniquely versatile and highly-attractive. 
%not based on a synthetic model of the instrument but on calibration with real data, makes it uniquely versatile and attractive.
Its has been successfully demonstrated in a variety of environments, both in the laboratory and on-sky, using different experimental platforms and coronagraphs. A common characteristics of these demonstrations is that the resulting DHs exhibit moderate to large inner working angles (IWAs) and wide fields of view (FoV), often extending over several tens of $\lambda/D$.
%In these tests, the resulting dark holes generally have large or moderate inner working angles (IWA) and wide field of view (FoV), spanning several tens of $\lambda/D$.

Here, we report the first demonstration of iEFC at small angular separations, achieving a DH spanning 1 to 4 $\lambda/D$ on a segmented pupil in H-band under laboratory conditions. This combination of aggressive IWA with reduced FoV highlights the uniqueness of the SPEED facility (segmented pupil experiment for exoplanet detection). SPEED is a high-contrast imaging testbed designed for extremely large telescopes (ELTs), developed in the context of the ELT/PCS instrument (planetary camera and spectrograph), currently in pre-phase A. Its small-IWA coronagraph, an APCMC (apodized pupil complex mask coronagraph), and its multi– and out-of–pupil-plane deformable mirrors (DMs) architecture are specifically optimized for DHs at very small IWAs, down to diffraction limit.

Our results demonstrate (i) that widely separated out-of-pupil DMs are well suited to generating DHs with IWAs down to 1$\lambda/D$, and (ii) the effectiveness of iEFC, when combined with the SPEED wavefront control architecture, for high-contrast imaging with segmented telescopes. Ultimately, these results pave the way toward meeting the stringent high-contrast requirements of ELT-class instruments, opening new opportunities for exoplanet detection at unprecedented angular resolutions.
%Our results demonstrate the potential of iEFC, when combined with SPEED’s WFC design choices, to ideally push the limits of high-contrast imaging down to $10^{-7}$ contrast in the dark hole. Ultimately, these results pave the road to meeting the stringent high-contrast requirements of ELT-class instruments, opening new opportunities for exoplanet detection at unprecedented angular resolutions.
\end{abstract}


%%%%%%%%%%%%%%%%%%%%%%%%%%%%%%%%%%%%%%%%%%%%%%%%%%%%%%%%%%%%%%%%
\section{Introduction}
\label{sec:Intro}

Direct imaging of exoplanets provides a unique opportunity to characterize planetary atmospheres and investigate the diversity of planetary systems beyond the Solar System. By directly separating the light emitted or reflected by a planet from that of its host star, this technique enables the measurement of spectroscopic signatures that can constrain planetary composition, temperature, and potential habitability. However, direct imaging remains one of the most challenging observational techniques in astronomy due to the extreme contrast and small angular separation between a planet and its host star. An Earth-like planet observed in reflected light around a Sun-like star can be several orders of magnitude fainter than its host star, while being separated by only a few tens of milliarcseconds.

The next generation of large astronomical observatories aims to overcome these limitations by combining large collecting areas with advanced high-contrast imaging techniques. %Future space-based missions dedicated to exoplanet characterization, as well as extremely large ground-based telescopes, rely on large primary mirrors to increase sensitivity and angular resolution. 
Although future space-based missions and extremely large ground-based telescopes (ELTs) address complementary scientific goals and exoplanet populations, they share common technological challenges. In particular, both rely on large primary mirrors to achieve the sensitivity and angular resolution required for high-contrast imaging.
Because of their unprecedented size, these telescopes require segmented primary
mirrors, introducing additional challenges for high-contrast imaging. %Discontinuities between segments, such as differential piston, tip-tilt, and higher-order phasing errors, create additional diffraction structures that degrade the achievable contrast. Controlling these aberrations is therefore essential to fully exploit the capabilities of segmented telescopes.
Coronagraphy and wavefront control (WFC) are key technologies developed to enable direct imaging of exoplanets. By modifying the optical propagation of the stellar wavefront, coronagraphs suppress the coherent starlight while preserving the off-axis planetary signal. However, the performance of a coronagraph is ultimately limited by residual wavefront aberrations originating from imperfect optics, alignment errors, atmospheric residuals, and telescope instabilities.
Consequently, active wavefront control is required to measure and correct these aberrations, and wavefront shaping is needed to create regions of high contrast, commonly referred to as dark holes, in the coronagraphic image. 

Several wavefront control techniques have been developed to achieve this goal. Electric field conjugation (EFC) relies on a calibrated model of the optical system to estimate the response of the focal-plane electric field to deformable mirror commands and compute the correction required to minimize residual starlight. Focal-plane wavefront sensing techniques based on pairwise probing (PWP) enable the estimation of the complex electric field by introducing known modulations of the wavefront. These approaches have demonstrated impressive contrast improvements on several high-contrast imaging testbeds\cite{Soummer_HICAT, Laginja_THD2} and instruments\cite{Potier_PWP_EFC_SPHERE}. However, their performance relies on the accuracy of the optical model and calibration, which can become challenging for complex optical systems.%, particularly for segmented telescopes with evolving aberration sources.
Implicit electric field conjugation (iEFC) provides an alternative, model-free approach to focal-plane wavefront control. Instead of explicitly estimating the complex electric field, iEFC directly calibrates the response of differential focal-plane intensity measurements to deformable mirror commands. The resulting interaction matrix allows the controller to determine corrections that minimize the measured residual intensity without requiring an explicit optical model. This approach is particularly attractive for complex optical systems where accurate modeling of all aberration sources is challenging.

In this work, we investigate the performance of iEFC on the segmented pupil experiment for exoplanet detection (SPEED) testbed at the Observatoire de la Côte d'Azur. SPEED is specifically designed to study high-contrast imaging for segmented telescope architectures and operates at extremely small inner working angles (IWAs), between 1 and 4 $\lambda/D$ in the H-band ($\lambda=1650$ nm). This angular separation range corresponds to approximately 8--35 mas for a 39-m class telescope, representative of the challenging regime required for future extremely large telescopes to detect Earth-like analogs around M-stars.
%The objective of this work is to demonstrate and characterize the capabilities of iEFC in this small inner working angle regime, using a segmented pupil representative of future large-aperture observatories. 
The objective of this work is to demonstrate the feasibility of high-contrast wavefront control at small IWAs using a segmented pupil representative of future large-aperture observatories. To this end, we generate a DH spanning 1 to 4$\lambda/D$ and evaluate the performance of iEFC in this previously unexplored regime.
The influence of the calibration procedure, probe generation, regularization strategy, and control parameters on the achievable contrast is explored, laying the groundwork for future studies on the application of model-free focal-plane wavefront control to segmented high-contrast imaging instruments.
%The impact of the calibration procedure, probe generation, regularization strategy, and control parameters on the achievable contrast is investigated, providing insights into the applicability of model-free focal-plane wavefront control for future segmented high-contrast imaging instruments.

%%%%%%%%%%%%%%%%%%%%%%%%%%%%%%%%%%%%%%%%%%%%%%%%%%%%%%%%%%%%%%%%
\section{The SPEED testbed}
\label{sec:SPEED}
% \subsection{Design and configuration}\label{subsec:SPEED context}

\begin{figure}
    \centering
    \begin{subfigure}[b]{0.58\textwidth}
        \centering
        \includegraphics[width=\textwidth]{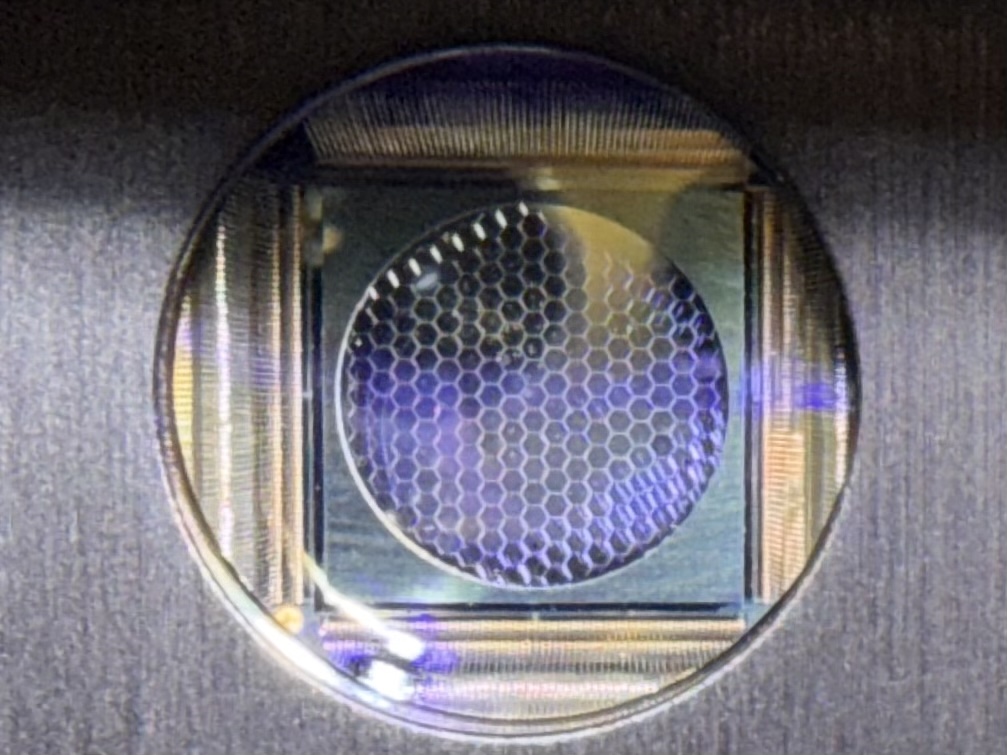}
        \caption{}
        \label{fig:DMs-a}
    \end{subfigure}
    \hfill
    \begin{subfigure}[b]{0.325\textwidth}
        \centering
        \includegraphics[width=\textwidth]{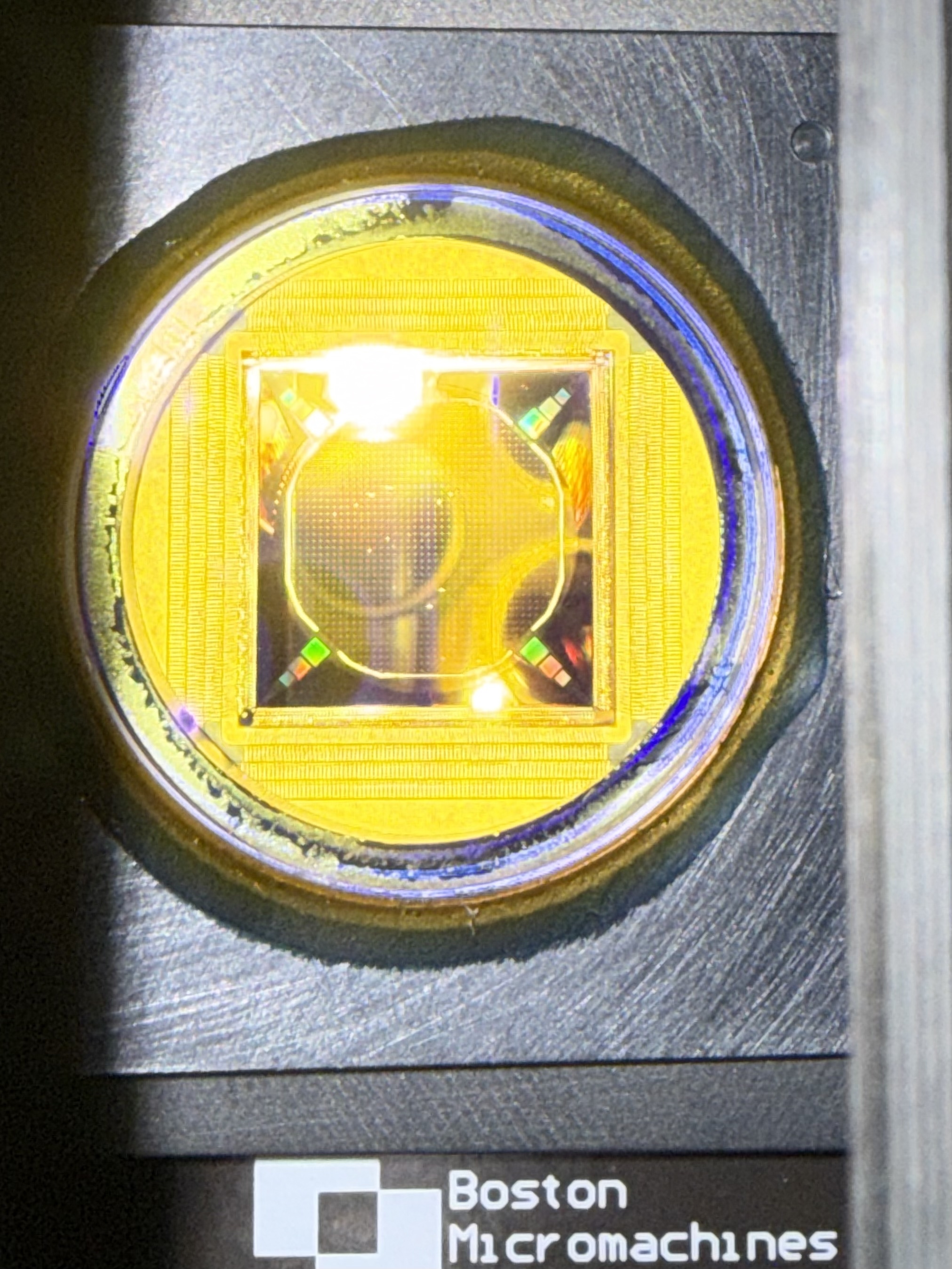}
        \caption{}
        \label{fig:DMs-b}
    \end{subfigure}
    \vspace{0.5em}

\caption{(a) IrisAO PTT489 segmented mirror comprising 163 actuators; (b) Boston Micromachines 34×34 Kilo-C deformable mirror.}

  %  \caption{(a) IrisAO PTT489 segmented mirror, with a total of 163 actuators; (b) Boston Micromachines $34\times34$ Kilo-C deformable mirror.}
    \label{fig:DMs}
\end{figure}

The segmented pupil experiment for exoplanet detection (SPEED) \cite{SPEED2014,SPEED2015} is a high-contrast imaging testbed located at the Observatoire de la Côte d’Azur, dedicated to developing and validating instrumental concepts capable of achieving deep contrast levels at very small angular separations for next-generation ground-based telescopes, particularly the European Extremely Large Telescope (ELT) and its future Planetary Camera and Spectrograph (PCS) instrument \cite{PCS}.
%The segmented pupil experiment for exoplanet detection (SPEED)\cite{SPEED2014,SPEED2015} is a high contrast imaging testbed located at the Côte d'Azur Observatory, aimed at coronagraphic development for the next generation ground-based telescopes, specifically the European extremely large telescope (ELT) and its last generation instrument PCS (planetary camera and spectrograph)\cite{PCS}.
To accurately reproduce the optical configuration of an ELT-like telescope, SPEED is equipped with a dedicated telescope emulator composed of an active segmented mirror (ASM; see Fig.~\ref{fig:DMs-a}) and a tip–tilt mirror onto which the entrance pupil mask is magnetically mounted. The ASM reproduces the main characteristics of the ELT primary mirror, including the segment arrangement, inter-segment discontinuities, and the central obstruction and spider geometry through the entrance pupil mask. Although the emulator uses a reduced number of segments (163 compared with the 798 segments of the ELT primary mirror), it preserves the essential diffraction properties of a segmented aperture, enabling realistic laboratory investigations of coronagraphic concepts and wavefront control strategies for ELT-class high-contrast imaging.

%To accurately emulate the ELT configuration, SPEED is equipped with a telescope emulator, consisting of an active segmented mirror (ASM - see fig.\ref{fig:DMs-a}), a tip-tilt mirror and an entrance pupil mask; those three elements are design following the ELT characteristics (inter-segment gap, spider location and size), with a reduced number of segments (163 instead of 798 for the ELT primary mirror).

\begin{figure}
    \centering
    \includegraphics[width=0.8\linewidth]{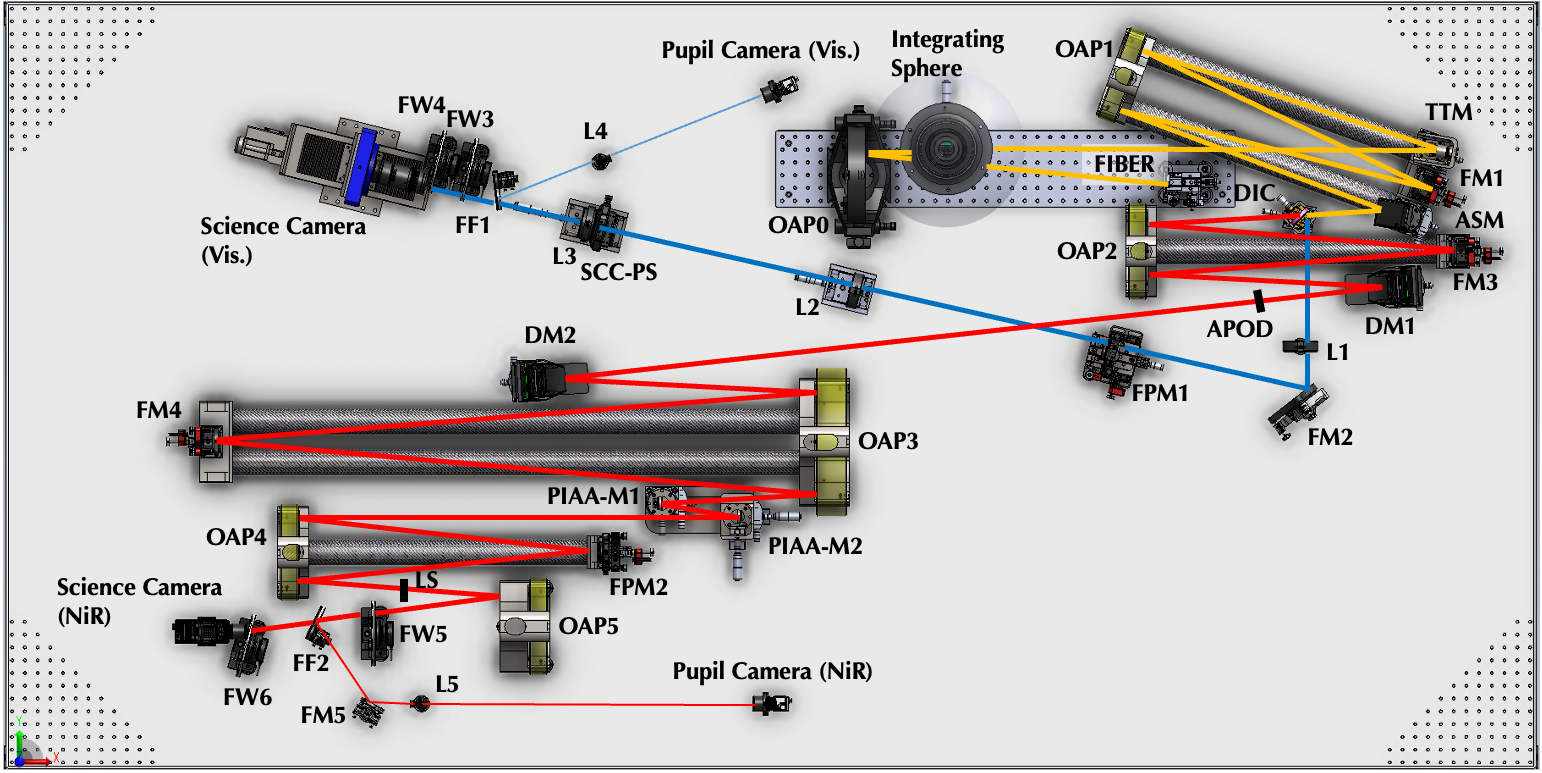}
    \caption{SPEED optical layout with the APCMC coronagraph. The apodizer is located in a pupil plane in between the two DMs. The PIAA-M1 \& M2 elements are replaced by flat mirrors in the APCMC configuration.}
    \label{fig:optical_layout}
\end{figure}

The SPEED testbed consists of two optical paths: a visible path ($\lambda = 632\ nm$) dedicated to segment cophasing, and an infrared path ($\lambda = 1650\ nm$) dedicated to high-contrast imaging. Wavefront shaping and control are achieved using two out-of-pupil deformable mirrors (DMs) in combination with a small-IWA coronagraphic module. The DMs are two 34×34 Kilo-C deformable mirrors manufactured by Boston Micromachines \cite{Bierden2011} (see Fig.~\ref{fig:DMs-b}).
The coronagraph implemented on SPEED is the apodized pupil complex mask coronagraph (APCMC) \cite{APCMC}. It retains the focal plane mask from a PIAACMC design, while replacing the PIAA module with an amplitude pupil-plane apodizer. The apodizer is positioned upstream, in a pupil plane located between the two DMs, while the PIAA mirrors (M1 and M2) are replaced by flat mirrors. The complete optical layout is presented in Fig.~\ref{fig:optical_layout}.
All the experiments reported in this proceeding were performed in the H-band using a narrow-band filter with a bandwidth of 19 nm ($\Delta \lambda / \lambda = 1\%$).

%The SPEED testbed consists of a visible path ($\lambda = 632\ nm$), used for cophasing and an infrared path ($\lambda = 1650\ nm$) for high-contrast imaging.
%The wavefront shaping and control are performed by two out-of-pupil deformable mirrors (DMs) and a coronagraph module. The deformable mirrors are two $34\times34$ Kilo-C DMs from Boston Micromachines\cite{Bierden2011}, see Fig.\ref{fig:DMs-b}.
%SPEED's coronagraph is the apodized pupil complex mask coronagraph (APCMC)\cite{APCMC}. It keeps the focal plane mask and Lyot stop from the previous PIAACMC, and trades the PIAA module for an apodizer. The apodizer is located upstream in a pupil plane in between the two DMs, while the PIAA mirrors (M1 and M2) are exchanged for flat mirrors. See fig.\ref{fig:optical_layout} for details.

%%%%%%%%%%%%%%%%%%%%%%%%%%%%%%%%%%%%%%%%%%%%%%%%%%%%%%%%%%%%%%%%
\section{implicit electric field conjugation}
\label{sec:iEFC theory}

The recently proposed iEFC \cite{iEFC_Haffert} is a model-free DH creation algorithm. It was introduced as an alternative to conventional EFC \cite{EFC_Giveon_2007}, which relies on an accurate model of the instrument to estimate the complex electric field and derive the appropriate corrective commands. In contrast, the iEFC paradigm does not require an explicit estimation or reconstruction of the electric field within the targeted DH. Instead, it uses differential images as the physical observable for wavefront control. This approach relies on the fact that these differential images are linearly related to the focal-plane electric field through the system response, and therefore contain sufficient information to drive the correction process. Moreover, while EFC is primarily a wavefront control technique that requires a separate wavefront sensing strategy, iEFC intrinsically combines wavefront sensing and control into a unified framework. This makes the method particularly attractive, in particular for complex high-contrast imaging systems such as SPEED, where building an end-to-end optical model with high-fidelity can be challenging.

\subsection{Theoretical background}

Although iEFC does not require an explicit electric field model during operation, its formulation originates from the linearized description of electric field propagation commonly used in EFC. Following the formulation introduced by Milani et al.~\cite{iEFC_Milani}, the incident pupil-plane electric field can be written as
\begin{equation}
    E_{init} = A(x,y)e^{i\Phi(x,y)},
\end{equation}
where $x$ and $y$ denote the coordinates in the pupil plane, $A(x,y)$ represents the pupil-plane amplitude aberrations and $\Phi(x,y)$ the phase aberrations. Because optical propagation and apodization act as linear operators on the electric field, the resulting focal-plane electric field can be expressed as the initial electric field propagated through a linear operator C that encapsulates the complete optical system, leading to
\begin{equation}
    E_{fin} = C\left\{ E_{init} \right\} = C\left\{A(x,y)e^{i\Phi(x,y)}\right\}.
\end{equation}

Since the optical system is composed of a series of alternating pupil and focal planes, the operator $C$ can be represented as a sequence of Fourier and inverse Fourier transforms describing the propagation of the wavefront through the optical system. Accounting for the effect of the DMs, where $\Phi_{\mathrm{DM}}(x,y)$ denotes the phase introduced by their surfaces, and assuming the small-phase approximation, the DM-induced phase can be written as
\begin{equation}
    e^{i\phi_{\mathrm{DM}}(x,y)} \approx 1+i\phi_{\mathrm{DM}}(x,y).
\end{equation}

The resulting focal-plane electric field can therefore be separated into the initial aberrated field and the perturbation introduced by the deformable mirror, where, under the Marechal approximation, that apply on all $\phi$ terms, we get
\begin{equation}
    E_{fin} \simeq C\left\{ A(x,y)e^{i\Phi(x,y)}\right\} + iC\left\{ \phi_{DM}(x,y)\right\}.
\end{equation}

In the following, bold symbols denote vectorized quantities. The focal-plane electric field is represented as a vector containing the complex electric field values at each controlled pixel, while DM commands are represented as vectors containing the actuator amplitudes. Using such vector representation, this expression becomes:
\begin{equation}
    C\left\{ A(x,y)e^{i\Phi(x,y)}\right\} = \mathbf{E_{ab}}, \qquad \qquad iC\left\{ \Phi_{DM}(x,y)\right\} = \mathbf{G_1 A_1},
\end{equation}

where $\mathbf{E_{ab}}$ denotes the electric field contributed the system aberrations and $\mathbf{G_1}$ the Jacobian matrix relating the DM actuator heights $\mathbf{A_1}$ to the corresponding electric field perturbation in the focal plane. The total electric field in the image plane can therefore be written as
\begin{equation}\label{eq:E_f in vect}
    \mathbf{E_{fin}} = \mathbf{E_{ab}} + G_1 \mathbf{A_1}.
\end{equation}

Extending this formulation to a two-DM configuration following the approach introduced by Give'on et al.~\cite{EFC_Giveon_2007}, Eq.~\ref{eq:E_f in vect} becomes
\begin{equation}\label{eq:E_f generalized}
    \mathbf{E_{fin}} = \mathbf{E_{ab}} + G_1 \mathbf{A_1} + G_2\mathbf{ A_2} = \mathbf{E_{ab}} + G_{iEFC}\mathbf{A},
\end{equation}
where $G_1$ and $G_2$ represent the response matrices of each deformable
mirror. Defining
\begin{equation}
    G = \begin{bmatrix}
    G_1 & G_2
    \end{bmatrix},
    \qquad 
    \mathbf{A} = \begin{bmatrix}
    \mathbf{A}_1\\
    \mathbf{A}_2
    \end{bmatrix},
\end{equation}

gives
\begin{equation}
    \mathbf{E_{fin}} = \mathbf{E}_{ab} + G\mathbf{A}.
\end{equation}

Since the control is performed in a modal basis rather than directly on individual DM actuators (using Fourier modes in this work), Eq.~\ref{eq:E_f generalized} can be reformulated using the modal coefficients $\mathbf{m_c}$. The selected modes, referred to as calibration modes, are stored in the matrix $M_{\mathrm{modes}}$. The dimensions of the resulting Jacobian matrix therefore depend on the number of modes included in the calibration process.

Finally we get
\begin{equation}\label{eq:E_f final}
    \mathbf{E_{fin}} = \mathbf{E_{ab}} + G_{iEFC}\mathbf{m_c}.
\end{equation}

\subsection{iEFC calibration}
The iEFC calibration consists of measuring how the differential focal-plane intensity varies in response to known DM perturbations. A calibration mode may correspond to a single actuator, a Fourier mode, or a Hadamard mode. The current implementation on SPEED uses a Fourier modal basis, although a Hadamard basis is also available and being investigated.

\begin{figure}
    \centering
    \includegraphics[width=1\linewidth]{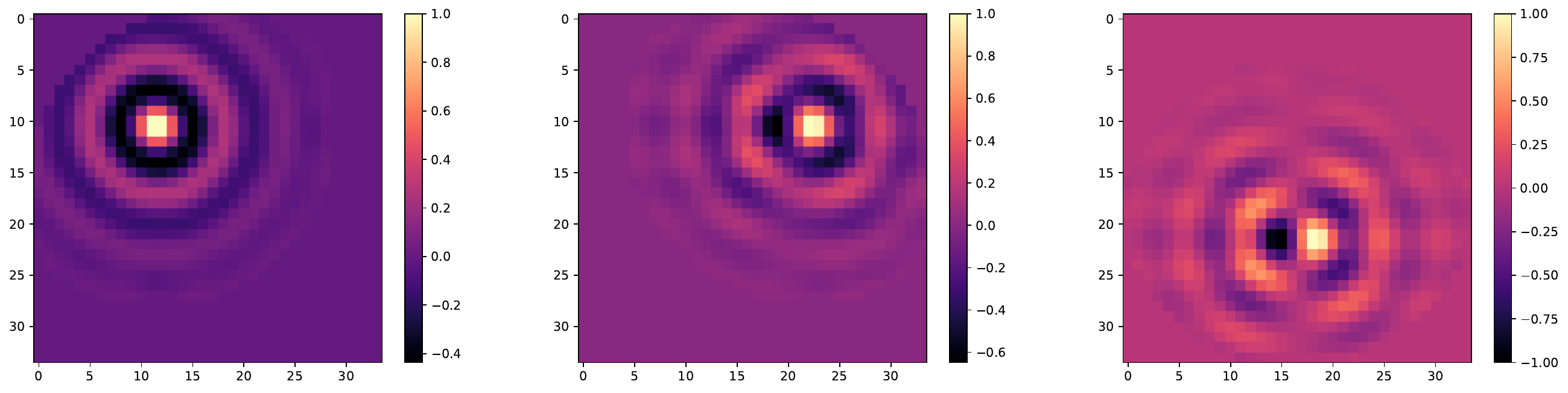}
    \caption{Fourier probes used for electric field estimation. Each probe command is represented on a $34\times34$ pixel grid, corresponding to the actuator layout of the Boston Micromachines DMs. The probe amplitudes are normalized, as their values are optimized during the iEFC control loop}
    \label{fig:probes}
\end{figure}

A set of Fourier DM probe shapes is used to modulate the focal-plane intensity during the calibration of each mode. Figure~\ref{fig:probes} shows the three Fourier DM probe shapes used throughout this work.
The DM probe shapes are generated using a Fourier modal basis selected to cover the spatial frequencies corresponding to the controlled region (i.e., the dark hole). The Fourier modes are sampled with a spatial-frequency interval of
\begin{equation}
\Delta f = 0.25~\lambda/D.
\end{equation}
%Let
%\begin{equation}
%\mathcal{F} = \left\{ (f_{x,i},f_{y,i}) \right\}_{i=1}^{N_f}
%\end{equation}
%denote the set of sampled spatial frequencies, where $N_f$ is the total number of Fourier frequencies. For each spatial frequency, $C_i(x,y)$ and $S_i(x,y)$ denote the corresponding cosine and sine Fourier modes defined on the DM actuator grid.
The set of sampled spatial frequencies is defined as
\begin{equation}
\mathcal{F} = \left\{ (f_{x,i},f_{y,i}) \right\}_{i=1}^{N_f},
\end{equation}
with $N_f$ being the total number of Fourier frequencies. For each frequency pair $(f_{x,i},f_{y,i})$, $C_i(x,y)$ and $S_i(x,y)$ denote the associated cosine and sine Fourier modes defined on the DM actuator grid.
To increase the contribution of high spatial frequencies, each Fourier mode is weighted according to its radial spatial frequency,
\begin{equation}
w_i = \frac{\sqrt{f_{x,i}^2+f_{y,i}^2}} {\displaystyle \max_j \sqrt{f_{x,j}^2+f_{y,j}^2}}.
\end{equation}

The weighted cosine and sine superpositions are then computed as
\begin{align}
C_{\mathrm{sum}}(x,y) &= \sum_{i=1}^{N_f} w_i\,C_i(x,y), \\
S_{\mathrm{sum}}(x,y) &= \sum_{i=1}^{N_f} w_i\,S_i(x,y).
\end{align}

%The DM probe shapes are obtained as linear combinations of these two superpositions,
The DM probe shapes, denoted $D_k$, are constructed from linear combinations of the cosine and sine Fourier mode superpositions defined above, such as
\begin{equation}
D_k(x,y) = c_k C_{\mathrm{sum}}(x,y) + s_k S_{\mathrm{sum}}(x,y),
\qquad
k=1,\ldots,N_{\mathrm{probe}},
\end{equation}
where the coefficients $(c_k,s_k)$ determine the relative contribution of the cosine and sine components, allowing the phase of each probe to be adjusted.
For the three probe shapes used in this work,
\begin{equation}
(c_k,s_k) = \left\{ (1,0), \left(\frac12,\frac12\right), (0,1) \right\}.
\end{equation}
This leads to the following probe functions:
\begin{align}
D_1 &= C_{\mathrm{sum}},\\
D_2 &= \frac12 C_{\mathrm{sum}}
      + \frac12 S_{\mathrm{sum}},\\
D_3 &= S_{\mathrm{sum}}.
\end{align}
Each DM probe shape is then mapped onto the actuator grid according to
\begin{equation}
\Delta_k = \left\{ (6,-5), (6,5), (-5,0) \right\},
\end{equation}
using bilinear interpolation while preserving the actuator grid dimensions and assigning zero values outside the original grid. The translated probe shapes are subsequently normalized,
\begin{equation}
\widetilde{D}_k(x,y) = \frac{D_k(x,y)} {\displaystyle \max_{x,y} D_k(x,y)},
\end{equation}
and multiplied by the binary DM actuator mask,
\begin{equation}
D_k^{\mathrm{DM}}(x,y) = \widetilde{D}_k(x,y) M_{\mathrm{DM}}(x,y),
\qquad
k=1,\ldots,N_{\mathrm{probe}},
\end{equation}
where $M_{\mathrm{DM}}$ denotes the binary mask defining the active DM actuators.

While the previous part described the generation of DM probes, the wavefront control process also requires an accurate calibration of the DM response. This calibration is performed using a set of predefined modes, referred to as calibration modes. For each calibration mode of amplitude $a_{\mathrm{mode}}$, the mode is first applied with a positive amplitude. Each DM probe shape is then successively added to and subtracted from the calibration mode with an amplitude $a_{\mathrm{probe}}$, producing the corresponding focal-plane intensity images $I_{+,i}$ and $I_{-,i}$ for the $i$-th probe shape. The procedure is subsequently repeated after applying the calibration mode with the opposite sign. For $N_{\mathrm{probe}}$ probe shapes, a total of $2N_{\mathrm{probe}}$ focal-plane images are therefore acquired for each sign of the calibration mode.
For the positive calibration mode, the differential image associated with the $i$-th probe shape is
\begin{equation}
\delta_{+,i} = \frac{I_{+,i}-I_{-,i}} {2a_{\mathrm{probe}}},
\end{equation}
while for the negative calibration mode,
\begin{equation}
\delta_{-,i} = \frac{I_{+,i}-I_{-,i}} {2a_{\mathrm{probe}}}.
\end{equation}
The differential measurements are stacked into the vectors:
\begin{equation}
\boldsymbol{\delta}_{+} = \begin{bmatrix}
\delta_{+,1}\\
\vdots\\
\delta_{+,N_{\mathrm{probe}}}
\end{bmatrix},
\qquad
\boldsymbol{\delta}_{-} = \begin{bmatrix}
\delta_{-,1}\\
\vdots\\
\delta_{-,N_{\mathrm{probe}}}
\end{bmatrix}.
\end{equation}
The response associated with calibration mode $m$ is then computed by finite differencing,
\begin{equation}
\mathbf{R}_{m,\delta} = \frac{\boldsymbol{\delta}_{+}-\boldsymbol{\delta}_{-}} {2a_{\mathrm{mode}}},
\end{equation}
where $\mathbf{R}_{m,\delta}$ represents the differential response of the focal-plane intensity to the considered calibration mode.
Repeating this procedure for all calibration modes yields the iEFC interaction matrix (or Jacobian),
\begin{equation}
G_{\mathrm{iEFC}} = \begin{bmatrix}
\mathbf{R}_{1,\delta} & \mathbf{R}_{2,\delta} & \cdots & \mathbf{R}_{N_{\mathrm{mode}},\delta} 
\end{bmatrix},
\end{equation}
whose columns correspond to the linear response of the differential focal-plane intensity measurements to each calibration mode. This interaction matrix is subsequently used by the iEFC controller to determine the DM correction that minimizes the measured differential intensity.

\subsection{iEFC Controller}
Once calibrated, the interaction matrix is inverted using a regularized least-squares approach to obtain the control matrix, which is then used to compute the DM corrections within the iEFC closed-loop control scheme.
%Once the interaction matrix has been calibrated, it is inverted using a regularized least-squares approach to obtain the control matrix governing the iEFC closed-loop correction. 
The iEFC controller operates directly on the differential focal-plane intensity measurements acquired with the DM probe shapes. At each iteration, the vector of differential measurements $\boldsymbol{\delta}$ is first acquired using the procedure described in the previous section. The corresponding modal correction is then obtained by applying the control matrix,
\begin{equation}
\mathbf{m_c} = -\mathbf{M}_{\mathrm{control}}\boldsymbol{\delta},
\end{equation}
where $\mathbf{m_c}$ contains the coefficients of the calibration modes used to update the DM. The associated DM command is subsequently reconstructed as
\begin{equation}
    \mathbf{A} = M_{\mathrm{modes}} \mathbf{m_c},
\end{equation}
where the columns of $M_{\mathrm{modes}}$ correspond to the calibration modal basis. The resulting command is applied to the DMs, and the procedure is repeated until the desired contrast is reached within the controlled DH.
The modal coefficients are obtained by minimizing the regularized least-squares cost function:
\begin{equation}
J = \left| \boldsymbol{\delta} + G_{\mathrm{iEFC}} \mathbf{m_c} \right|_2^2 + \lambda \left|  \mathbf{m_c} \right|_2^2, 
\end{equation}
where the first term minimizes the residual differential intensity while the second penalizes large modal amplitudes, using a regularization parameter (denoted, $\lambda$). Rather than using a fixed Tikhonov regularization parameter, the controller employs the $\beta$-regularization introduced by Sidick et al.~\cite{Sidick_beta_reg}, for which the control matrix is given by
\begin{equation}
\mathbf{M}_{\mathrm{control}} = \left(G_{\mathrm{iEFC}}^\top G_{\mathrm{iEFC}} + \alpha^2 10^\beta I\right)^{-1} G_{\mathrm{iEFC}}^\top,
\end{equation}
with
\begin{equation}
\alpha^2 = \max \left(\mathrm{diag} \left(G_{\mathrm{iEFC}}^\top G_{\mathrm{iEFC}} \right)\right),
\end{equation}
where $I$ denotes the identity matrix. The parameter $\beta$ controls the strength of the regularization: increasing $\beta$ produces more conservative corrections with smaller modal amplitudes, whereas decreasing $\beta$ allows more aggressive corrections at the expense of increased sensitivity to model errors and measurement noise.

%%%%%%%%%%%%%%%%%%%%%%%%%%%%%%%%%%%%%%%%%%%%%%%%%%%%%%%%%%%%%%%%
\section{Experimental results}
\label{sec:Lab results}

\subsection{Experimental conditions and data reduction}

All experiments presented in this work were performed on SPEED using a narrow-band source centered at a wavelength of $1650~\mathrm{nm}$, with a spectral bandwidth of $19~\mathrm{nm}$ ($\Delta\lambda/\lambda_0 \approx 1\%$). Unless otherwise specified, all DHs presented in the following sections were obtained using the same iEFC configuration and acquisition parameters.
The focal-plane images used by the controller were acquired with an exposure time of $20~\mathrm{ms}$ per frame. Each measurement used by the control loop corresponds to the average of 10 consecutive frames (NDIT = 10), reducing the impact of temporal fluctuations and detector noise. The correction commands were applied using a loop gain of $0.4$, meaning that only 40$\%$ of the computed correction was applied at each iteration in order to improve loop stability. 

%The iEFC interaction matrix was calibrated using a Fourier modal basis. For the two-DM configuration of SPEED, a total of 656 calibration modes were used, corresponding to 328 modes per DM. 
The iEFC interaction matrix was calibrated using a Fourier modal basis. In the two-DM configuration of SPEED, the calibration was performed using 656 modes in total, equally distributed between the two DMs (328 modes per DM).
Each DM basis therefore contained 164 cosine and 164 sine Fourier modes. The amplitude of the calibration modes was set to $30~\mathrm{nm}$ during both the interaction matrix acquisition and the control process. The same amplitude was used for the DM probes employed during the differential intensity measurements.
The iEFC control matrix was computed using $\beta$-regularization with a constant value of $\beta=-1.5$ for all experiments unless explicitly stated otherwise. The evolution of the DHs presented in the following sections therefore reflects changes in the DH geometry or control strategy, rather than variations in the acquisition parameters.

The DH contrast reported throughout this work is computed from the normalized intensity within the selected DH region. After dark subtraction and normalization by the maximum intensity of the off-axis PSF, the pixels contained within the DH region are extracted and averaged. Because the subtraction process can lead to small negative intensity values due to noise fluctuations, only positive pixels are considered in the contrast calculation. This choice avoids artificially decreasing the measured contrast and provides a conservative estimate of the achieved starlight suppression. Although this metric does not provide a complete statistical characterization of the DH performance, it offers a robust and consistent measurement of the contrast level obtained during the experiments.

\subsection{Impact of size and shape}
All DHs generated in this study were full or half-annuli. The ultimate objective was to demonstrate the creation of an annular DH extending from 1 to 4 $\lambda/D$, corresponding to the small IWA regime targeted by SPEED. Since creating and maintaining a DH close to the coronagraphic PSF core is intrinsically more challenging, we first investigated regions located at larger angular separations, where the residual stellar intensity is lower and the requirements on wavefront correction are less demanding. 

\begin{figure}
    \centering
    \includegraphics[width=0.8\linewidth]{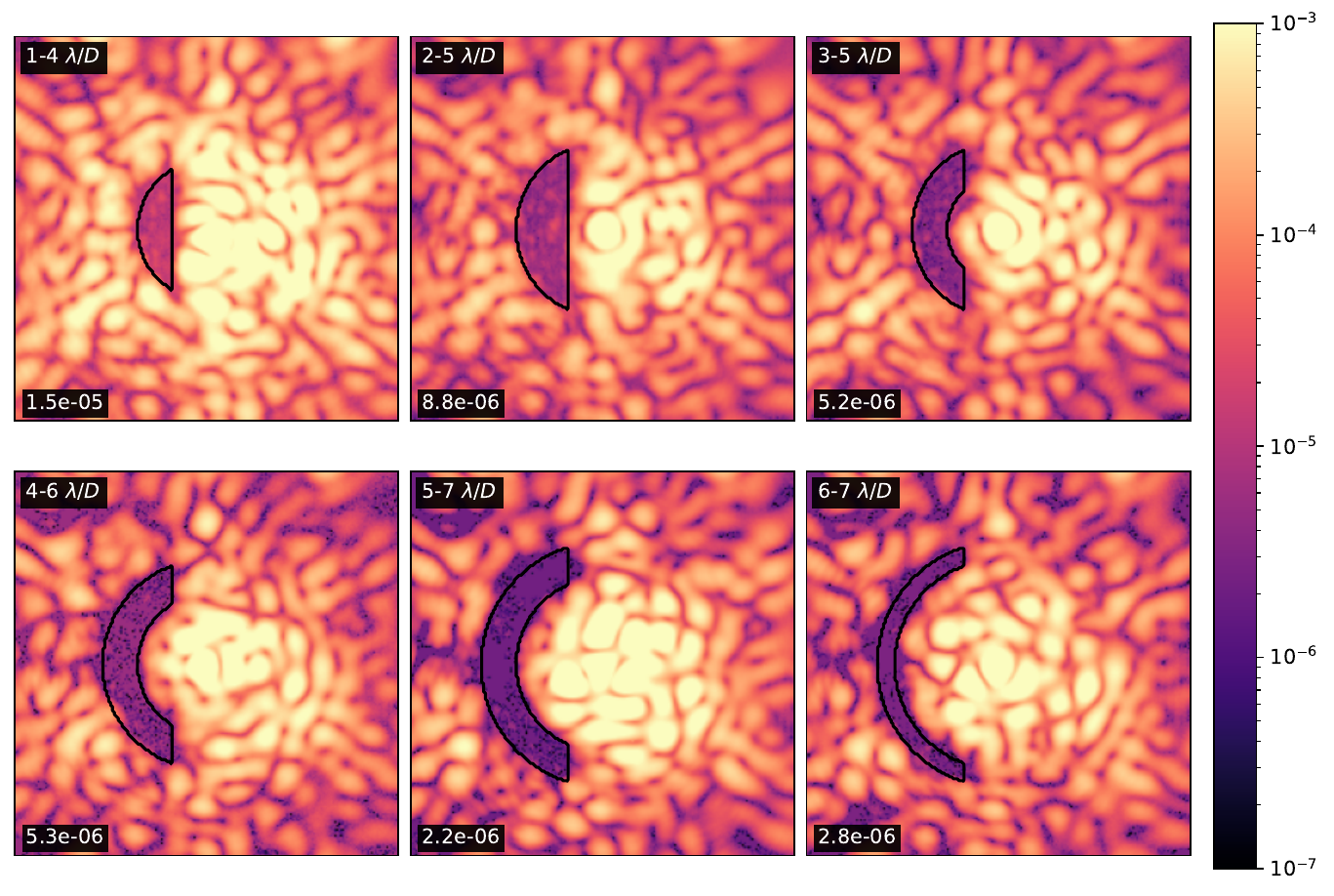}
    \caption{Different DH configurations tested on SPEED. The IWA ranges from 1 to 6 $\lambda/D$, and the outer working angle (OWA) from 4 to 7 $\lambda/D$. The reported contrast corresponds to the mean intensity within the DH region normalized by the maximum intensity of the off-axis PSF.}
    \label{fig:dh shape and size}
\end{figure}

Figure~\ref{fig:dh shape and size} summarizes the different DH geometries tested on SPEED. As expected, DHs located at larger angular separations achieve improved contrast performance, as the coronagraphic residual intensity decreases with increasing separation from the optical axis. Conversely, regions closer to the PSF core are more strongly affected by low-order aberrations, coronagraph leakage, and calibration errors, making the creation and maintenance of a stable DH more challenging at this stage. 
%The first configurations investigated were annular regions extending from 5--7 $\lambda/D$, followed by 4--6 $\lambda/D$, 3--5 $\lambda/D$, and finally the targeted 1--4 $\lambda/D$ region. 
The DH optimization was first performed at moderate IWAs, using annular regions spanning 5--7 $\lambda/D$, 4--6 $\lambda/D$, and 3--5 $\lambda/D$, before addressing the final target region of 1--4 $\lambda/D$.
Contrasts better than $3\times10^{-6}$ were obtained for DHs located beyond 5 $\lambda/D$. However, achieving stable convergence below the $10^{-5}$ contrast level remained challenging for the 1--4 $\lambda/D$ DH.

Due to the difficulty of reliably closing the control loop at such a small IWA, the final configuration was adjusted to a 2--5 $\lambda/D$ DH. This region remains close enough to the PSF core to be representative of the small-IWA regime targeted by future high-contrast instruments, while providing sufficient stability for a reproducible wavefront control demonstration. With this configuration, a contrast of $8\times10^{-6}$ was achieved in a clear and homogeneous dark region, demonstrating the potential of iEFC for small-IWA correction on SPEED.
%The contrast values reported in this section are computed using the metric described in Section~\ref{sec:experimental_conditions}.
These results represent a first assessment of the performance achievable with the current calibration and control strategy. Further work is needed to optimize the calibration process, probe generation, regularization, and control parameters before evaluating the ultimate contrast performance and identifying the fundamental limitations of the approach.

\subsection{Impact of regularization}

\begin{figure}[htbp]
    \centering

    \begin{subfigure}[t]{0.49\textwidth}
        \centering
        \includegraphics[width=\linewidth]{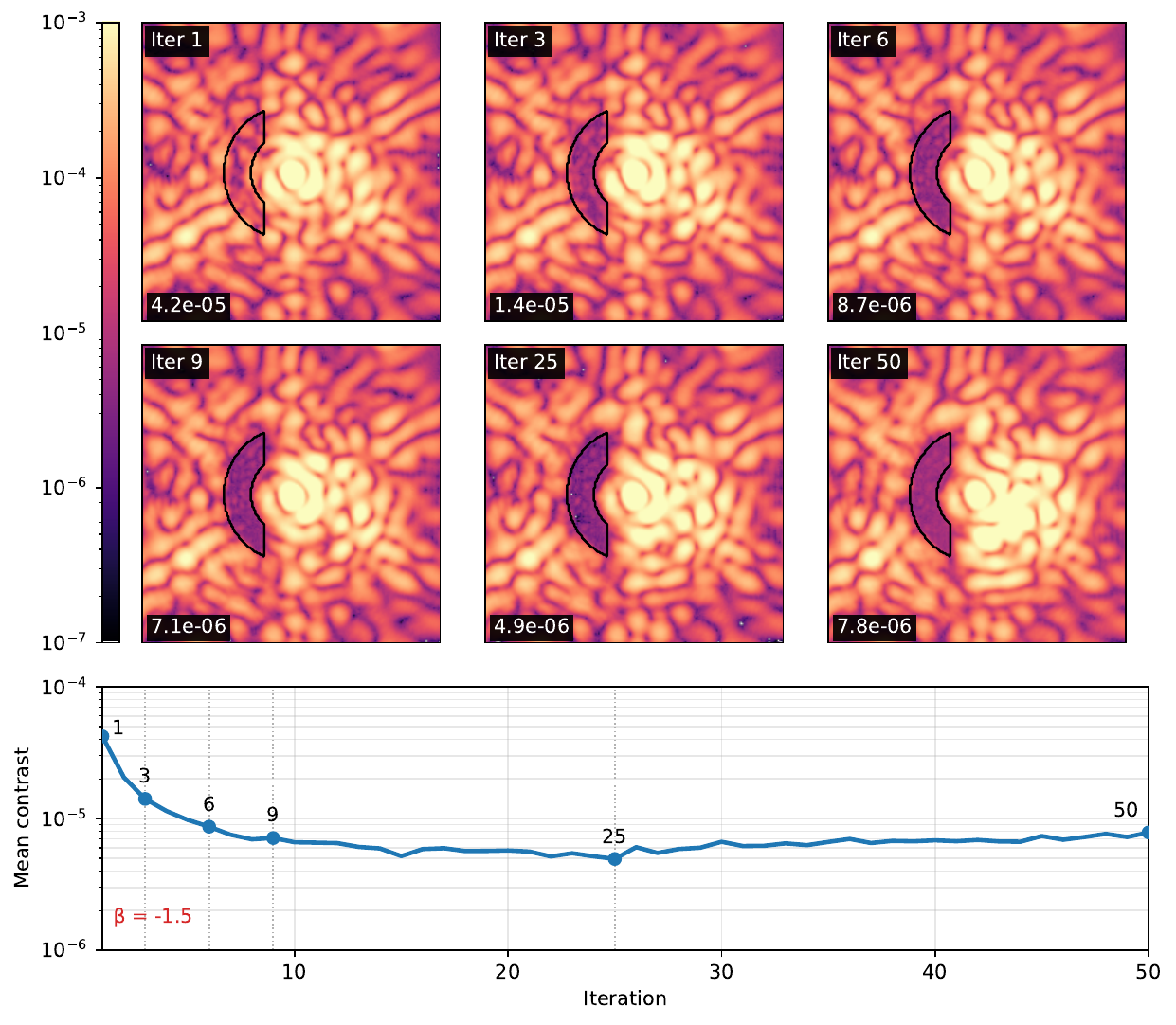}
        \caption{}
        \label{fig:DH-evl-over-iterations-with-fixed-beta-reg}
    \end{subfigure}
    \hfill
    \begin{subfigure}[t]{0.49\textwidth}
        \centering
        \includegraphics[width=\linewidth]{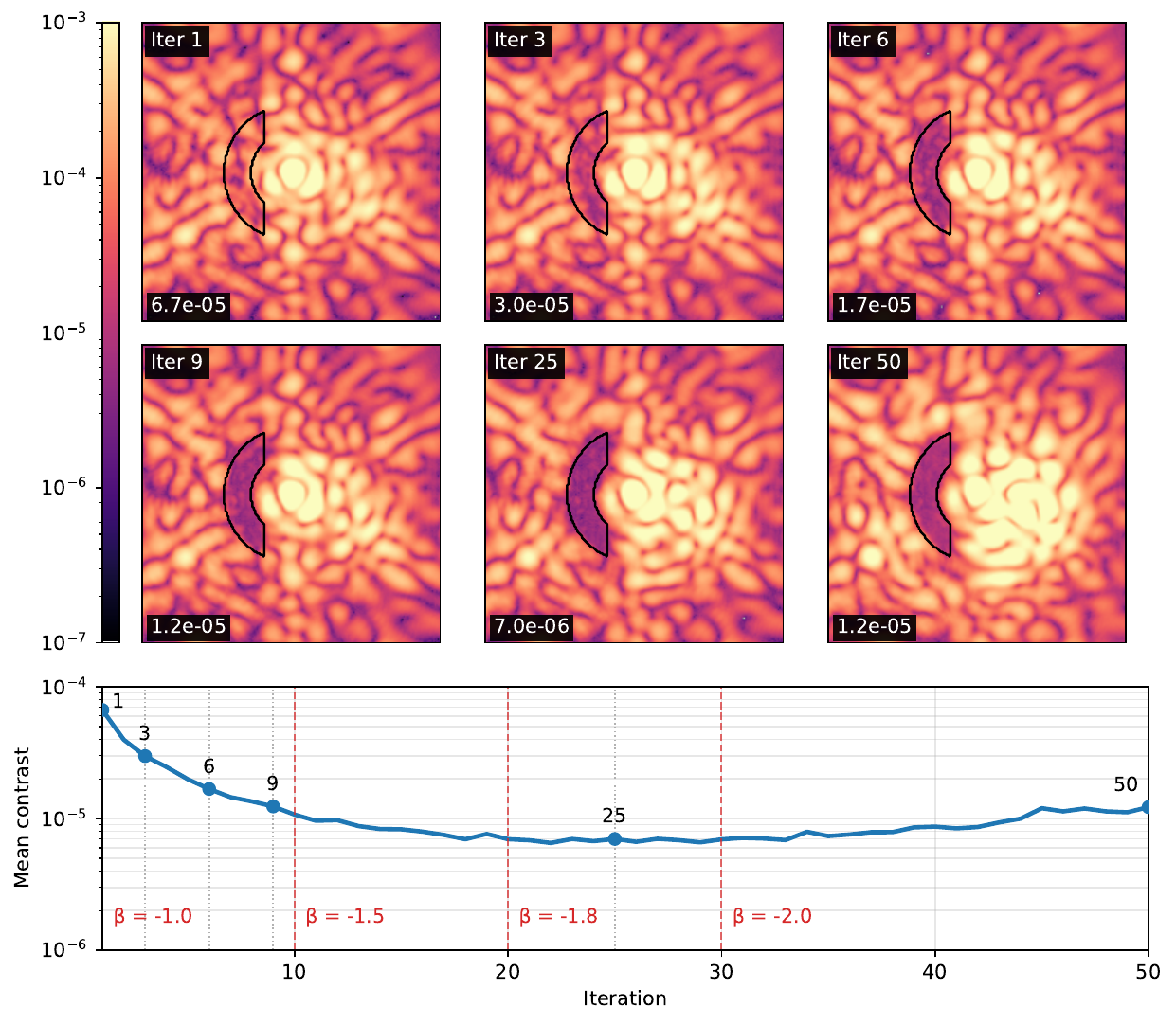}
        \caption{}
        \label{fig:DH-evl-over-iterations-with-variable-beta-reg}
    \end{subfigure}

    \caption{Evolution of a $3-5~\lambda/D$ DH over 50 iterations of the iEFC controller for two runs with quasi-identical configurations. The probe amplitude is $30~\mathrm{nm}$ and the spectral bandwidth is $19~\mathrm{nm}$ ($1\%$ at $1650~\mathrm{nm}$). In Fig.~\ref{fig:DH-evl-over-iterations-with-fixed-beta-reg}, the $\beta$-regularization parameter is kept constant at $\beta=-1.5$. In Fig.~\ref{fig:DH-evl-over-iterations-with-variable-beta-reg}, the $\beta$-regularization parameter is modified every 10 iterations.}
    \label{fig:beta-reg-impact}
\end{figure}

The impact of the regularization strategy on the iEFC performance was investigated using a $3-5~\lambda/D$ DH. Figure~\ref{fig:beta-reg-impact} shows the evolution of the contrast over 50 iterations of the iEFC controller for two runs with quasi-identical configurations. In both cases, a plateau is reached after approximately 10 iterations, with a contrast level of $\sim7\times10^{-6}$. For the run using a fixed regularization parameter ($\beta=-1.5$), the best performance is obtained around the 25th iteration, reaching a contrast of $4.9\times10^{-6}$.

The $\beta$-regularization parameter controls the trade-off between correction strength and robustness against calibration errors and noise. Lower values of $\beta$ reduce the amount of regularization, allowing larger corrections but increasing the sensitivity of the controller to calibration uncertainties, measurement noise, and non-linear effects in the DM response. Conversely, larger values of $\beta$ produce more conservative corrections with reduced modal amplitudes.

A $\beta$-scheduling strategy, where the regularization parameter is gradually modified throughout the control loop, was investigated following the approach described by Sidick et al.~\cite{Sidick_beta_reg}. It was expected that reducing the regularization during the loop would allow progressively stronger corrections and produce visible changes in the contrast evolution. However, no significant improvement or characteristic step-like behavior was observed following changes in $\beta$.
The absence of a clear dependence on the regularization schedule suggests that the achieved contrast level is not yet strongly driven by the regularization strength. Instead, the current performance is likely dominated by other error sources, such as uncertainties in the experimentally calibrated interaction matrix, measurement noise, temporal instabilities, residual aberrations, and nonlinearities in the DM response. Further optimization and characterization of these effects will be required to assess the ultimate contrast limitations of the control approach.
\section{Conclusion}
\label{sec:Conclusion}

The experiment presented in this work provides the first demonstration of a DH with our two out-of-pupil DM architecture, initially motivated by the theoretical and numerical studies presented in Beaulieu et al.~\cite{BEAULIEU2017,BEAULIEU2020}. The results obtained on SPEED provide an experimental demonstration of the potential of this architecture for high-contrast wavefront control in the context of a segmented pupil configuration representative of future large-aperture observatories.
We also provide an experimental demonstration of iEFC for high-contrast wavefront control on a segmented pupil testbed operating in the small IWA regime. Using a two out-of-pupil deformable mirror configuration and a Fourier modal basis, we achieved mean DH contrasts at the few $10^{-6}$ level, with a stable and homogeneous correction region in the $2-5~\lambda/D$ range. This demonstrates the capability of iEFC to perform focal-plane wavefront sensing and control at angular separations relevant for future extremely large telescopes.

The DH geometry was progressively investigated, starting from larger angular separations and moving toward the targeted small-IWA regime. While DHs located farther from the coronagraphic PSF core enabled deeper contrast levels, these experiments also indicate that the current performance is likely not limited by the intrinsic capabilities of the control algorithm, but instead by experimental limitations and by the need for further optimization of the calibration and control strategies. In particular, the transition toward smaller IWAs, as targeted by the SPEED project, will require further investigations for identifying and mitigating the contrast limitations of the current setup to reach deeper contrast levels. In particular, improvements in calibration procedures, probe optimization, and control strategies will be investigated. Additional studies will also be required to evaluate the impact of broader spectral bandwidths and to assess the long-term stability of the correction. %allowed the operational boundaries of the current configuration to be explored and identified the $2$--$5~\lambda/D$ region as a suitable compromise between achievable contrast and control-loop robustness.

%The two out-of-pupil deformable mirror architecture investigated in this work was initially motivated by the simulations presented in Beaulieu et al.~\cite{BEAULIEU2017,BEAULIEU2020}. The results obtained on SPEED provide an experimental validation of this architecture for high-contrast wavefront control and extend previous numerical studies toward a segmented pupil configuration representative of future large-aperture observatories. 

%Further work will focus on identifying and mitigating the contrast limitations of the current setup to reach deeper contrast levels. In particular, improvements in calibration procedures, temporal stability, probe optimization, and control strategies will be investigated. Additional studies will also be required to evaluate the impact of broader spectral bandwidths and to assess the long-term stability of the correction.

%The combination of the SPEED segmented-pupil architecture and the APCMC coronagraph provides a unique platform for studying wavefront sensing and control techniques required for future segmented telescopes. The results presented in this work demonstrate the potential of model-free focal-plane control approaches such as iEFC for addressing the challenges associated with small inner working angle high-contrast imaging and represent a step toward their application on future exoplanet imaging facilities.

%%%%%%%%%%%%%%%%%%%%%%%%%%%%%%%%%%%%%%%%%%%%%%%%%%%%%%%%%%%%%%%%
\acknowledgments The activity outlined in this paper is partially funded by the French Government, and by the European Union as part of the FEDER program, the PACA Region, the Observatoire de la Côte d’Azur, the Lagrange Laboratory, the Université de la Côte d’Azur, Airbus Defence and Space, the French spatial agency (CNES), Thales Alenia Space and the Action Spécifique Haute Résolution Angulaire (ASHRA) of CNRS/INSU co-funded by CNES. In addition, the activity concerning the development of the APCMC received funding from CNES under contracts 4500063765/DIA094, 4500049994/DIA094 and R-S23/SU-0002-088-03. C. S. acknowledges support from Thales Alenia Space and Région PACA. The authors sincerely thanks R. Soummer and I. Laginja for insightful discussions and valuable advice on DH testing. 

%%%%%%%%%%%%%%%%%%%%%%%%%%%%%%%%%%%%%%%%%%%%%%%%%%%%%%%%%%%%%%%%
\bibliography{biblio} % bibliography data in report.bib

@inproceedings{APCMC,
  adsurl    = {https://ui.adsabs.harvard.edu/abs/2024SPIE13097E..6RS},
  author    = {{Sallard}, C. and {Martinez}, P. and {Spang}, A. and {Marcotto}, A. and {Beaulieu}, M. and {Gouvret}, C. and {Dejonghe}, J.},
  booktitle = {Adaptive Optics Systems IX},
  doi       = {10.1117/12.3019025},
  editor    = {{Jackson}, Kathryn J. and {Schmidt}, Dirk and {Vernet}, Elise},
  eid       = {130976R},
  month     = aug,
  pages     = {130976R},
  series    = {Society of Photo-Optical Instrumentation Engineers (SPIE) Conference Series},
  title     = {{Phase-induced amplitude apodization complex mask coronagraph (PIAACMC) without PIAA: redesigning a phase-induced amplitude apodization to a conventional pupil amplitude apodization}},
  volume    = {13097},
  year      = 2024
}

@article{BEAULIEU2017,
  adsurl   = {http://cdsads.u-strasbg.fr/abs/2017MNRAS.469..218B},
  author   = {{Beaulieu}, M. and {Abe}, L. and {Martinez}, P. and {Baudoz}, P. and 
              {Gouvret}, C. and {Vakili}, F.},
  doi      = {10.1093/mnras/stx636},
  journal  = {\mnras},
  month    = jul,
  pages    = {218-230},
  title    = {{High-contrast imaging at small separations: impact of the optical configuration of two deformable mirrors on dark holes}},
  volume   = 469,
  year     = 2017
}

@article{BEAULIEU2020,
  author   = {Beaulieu, M and Martinez, P and Abe, L and Gouvret, C and Baudoz, P and Galicher, R},
  doi      = {10.1093/mnras/staa2106},
  eprint   = {https://academic.oup.com/mnras/article-pdf/498/3/3914/33782741/staa2106.pdf},
  issn     = {0035-8711},
  journal  = {Monthly Notices of the Royal Astronomical Society},
  month    = {07},
  number   = {3},
  pages    = {3914-3926},
  title    = {{High contrast at small separation – II. Impact on the dark hole of a realistic optical set-up with two deformable mirrors}},
  url      = {https://doi.org/10.1093/mnras/staa2106},
  volume   = {498},
  year     = {2020}
}

@inproceedings{Bierden2011,
  adsurl    = {https://ui.adsabs.harvard.edu/abs/2011aoel.confE..34B},
  author    = {{Bierden}, Paul and {Cornelissen}, Steven and {Lam}, Charlie and {Bifano}, Thomas},
  booktitle = {Second International Conference on Adaptive Optics for Extremely Large Telescopes. Online at <A href=``http://ao4elt2.lesia.obspm.fr''>http://ao4elt2.lesia.obspm.fr</A},
  eid       = {34},
  month     = sep,
  pages     = {34},
  title     = {{MEMS Deformable Mirrors in Astronomical AO}},
  year      = 2011
}

@inproceedings{EFC_Giveon_2007,
  adsurl    = {https://ui.adsabs.harvard.edu/abs/2007SPIE.6691E..0AG},
  author    = {{Give'on}, Amir and {Kern}, Brian and {Shaklan}, Stuart and {Moody}, Dwight C. and {Pueyo}, Laurent},
  booktitle = {Astronomical Adaptive Optics Systems and Applications III},
  doi       = {10.1117/12.733122},
  editor    = {{Tyson}, Robert K. and {Lloyd-Hart}, Michael},
  eid       = {66910A},
  month     = sep,
  pages     = {66910A},
  series    = {Society of Photo-Optical Instrumentation Engineers (SPIE) Conference Series},
  title     = {{Broadband wavefront correction algorithm for high-contrast imaging systems}},
  volume    = {6691},
  year      = 2007
}

@article{iEFC_Haffert,
  adsurl        = {https://ui.adsabs.harvard.edu/abs/2023A&A...673A..28H},
  archiveprefix = {arXiv},
  author        = {{Haffert}, S.~Y. and {Males}, J.~R. and {Ahn}, K. and {Van Gorkom}, K. and {Guyon}, O. and {Close}, L.~M. and {Long}, J.~D. and {Hedglen}, A.~D. and {Schatz}, L. and {Kautz}, M. and {Lumbres}, J. and {Rodack}, A. and {Knight}, J.~M. and {Miller}, K.},
  doi           = {10.1051/0004-6361/202244960},
  eid           = {A28},
  eprint        = {2303.13719},
  journal       = {\aap},
  month         = may,
  pages         = {A28},
  primaryclass  = {astro-ph.IM},
  title         = {{Implicit electric field conjugation: Data-driven focal plane control}},
  volume        = {673},
  year          = 2023
}

@article{iEFC_Milani,
  adsurl        = {https://ui.adsabs.harvard.edu/abs/2024JATIS..10b9001M},
  archiveprefix = {arXiv},
  author        = {{Milani}, Kian and {Douglas}, Ewan S. and {Haffert}, Sebastiaan Y. and {Van Gorkom}, Kyle},
  doi           = {10.1117/1.JATIS.10.2.029001},
  eid           = {029001},
  eprint        = {2405.03899},
  journal       = {Journal of Astronomical Telescopes, Instruments, and Systems},
  month         = apr,
  pages         = {029001},
  primaryclass  = {astro-ph.IM},
  title         = {{Modeling and performance analysis of implicit electric field conjugation with two deformable mirrors applied to the Roman Coronagraph}},
  volume        = {10},
  year          = 2024
}

@article{Laginja_THD2,
  adsurl        = {https://ui.adsabs.harvard.edu/abs/2025A&A...698A.130L},
  archiveprefix = {arXiv},
  author        = {{Laginja}, Iva and {Baudoz}, Pierre and {Mazoyer}, Johan and {Potier}, Axel and {Galicher}, Rapha{\"e}l and {Boussaha}, Faouzi},
  doi           = {10.1051/0004-6361/202553797},
  eid           = {A130},
  eprint        = {2504.01064},
  journal       = {\aap},
  month         = jun,
  pages         = {A130},
  primaryclass  = {astro-ph.IM},
  title         = {{Extended linearity in the high-order wavefront sensor for the Roman Coronagraph}},
  volume        = {698},
  year          = 2025
}

@article{PCS,
  adsurl        = {https://ui.adsabs.harvard.edu/abs/2021Msngr.182...38K},
  archiveprefix = {arXiv},
  author        = {{Kasper}, M. and {Cerpa Urra}, N. and {Pathak}, P. and {Bonse}, M. and {Nousiainen}, J. and {Engler}, B. and {Heritier}, C.~T. and {Kammerer}, J. and {Leveratto}, S. and {Rajani}, C. and {Bristow}, P. and {Le Louarn}, M. and {Madec}, P. -Y. and {Str{\"o}bele}, S. and {Verinaud}, C. and {Glauser}, A. and {Quanz}, S.~P. and {Helin}, T. and {Keller}, C. and {Snik}, F. and {Boccaletti}, A. and {Chauvin}, G. and {Mouillet}, D. and {Kulcs{\'a}r}, C. and {Raynaud}, H. -F.},
  doi           = {10.18727/0722-6691/5221},
  eprint        = {2103.11196},
  journal       = {The Messenger},
  month         = mar,
  pages         = {38-43},
  primaryclass  = {astro-ph.IM},
  title         = {{PCS {\textemdash} A Roadmap for Exoearth Imaging with the ELT}},
  volume        = {182},
  year          = 2021
}

@article{Potier_PWP_EFC_SPHERE,
  author  = {{Potier, A.} and {Galicher, R.} and {Baudoz, P.} and {Huby, E.} and {Milli, J.} and {Wahhaj, Z.} and {Boccaletti, A.} and {Vigan, A.} and {N’Diaye, M.} and {Sauvage, J.-F.}},
  doi     = {10.1051/0004-6361/202038010},
  journal = {A\&A},
  pages   = {A117},
  title   = {Increasing the raw contrast of VLT/SPHERE with the dark hole technique - I. Simulations and validation on the internal source},
  url     = {https://doi.org/10.1051/0004-6361/202038010},
  volume  = 638,
  year    = 2020
}

@inproceedings{Sidick_beta_reg,
  adsurl    = {https://ui.adsabs.harvard.edu/abs/2017SPIE10400E..22S},
  author    = {{Sidick}, Erkin and {Seo}, Byoung-Joon and {Kern}, Brian and {Marx}, David and {Poberezhskiy}, Ilya and {Nemati}, Bijan},
  booktitle = {Society of Photo-Optical Instrumentation Engineers (SPIE) Conference Series},
  doi       = {10.1117/12.2274440},
  editor    = {{Shaklan}, Stuart},
  eid       = {1040022},
  month     = sep,
  pages     = {1040022},
  series    = {Society of Photo-Optical Instrumentation Engineers (SPIE) Conference Series},
  title     = {{Optimizing the regularization in broadband wavefront control algorithm for WFIRST coronagraph}},
  volume    = {10400},
  year      = 2017
}

@inproceedings{Soummer_HICAT,
       author = {{Soummer}, R{\'e}mi and {Por}, Emiel H. and {Pourcelot}, Rapha{\"e}l. and {Redmond}, Susan and {Laginja}, Iva and {Will}, Scott D. and {Perrin}, Marshall D. and {Pueyo}, Laurent and {Sahoo}, Ananya and {Petrone}, Peter and {Brooks}, Keira J. and {Fox}, Rachel and {Klein}, Alex and {Nickson}, Bryony and {Comeau}, Thomas and {Ferrari}, Marc and {Gontrum}, Rob and {Hagopian}, John and {Leboulleux}, Lucie and {Leongomez}, Daniel and {Lugten}, Joe and {Mugnier}, Laurent M. and {N'Diaye}, Mamadou and {Nguyen}, Meiji and {Noss}, James and {Sauvage}, Jean-Fran{\c{c}}ois and {Scott}, Nathan and {Sivaramakrishnan}, Anand and {Subedi}, Hari B. and {Weinstock}, Sam},
        title = "{High-contrast imager for complex aperture telescopes (HiCAT): 8. Dark zone demonstration with simultaneous closed-loop low-order wavefront sensing and control}",
    booktitle = {Space Telescopes and Instrumentation 2022: Optical, Infrared, and Millimeter Wave},
         year = 2022,
       editor = {{Coyle}, Laura E. and {Matsuura}, Shuji and {Perrin}, Marshall D.},
       series = {Society of Photo-Optical Instrumentation Engineers (SPIE) Conference Series},
       volume = {12180},
        month = aug,
          eid = {1218026},
        pages = {1218026},
          doi = {10.1117/12.2630444},
archivePrefix = {arXiv},
       eprint = {2409.13026},
 primaryClass = {astro-ph.IM},
       adsurl = {https://ui.adsabs.harvard.edu/abs/2022SPIE12180E..26S}
}

@inproceedings{SPEED2014,
  adsurl    = {http://adsabs.harvard.edu/abs/2014SPIE.9145E..4EM},
  author    = {{Martinez}, P. and {Preis}, O. and {Gouvret}, C. and {Dejonghe}, J. and 
               {Daban}, J.-B. and {Spang}, A. and {Martinache}, F. and {Beaulieu}, M. and 
               {Janin-Potiron}, P. and {Abe}, L. and {Fantei-Caujolle}, Y. and 
               {Mattei}, D. and {Ottogalli}, S.},
  booktitle = {Ground-based and Airborne Telescopes V},
  doi       = {10.1117/12.2055338},
  eid       = {91454E},
  month     = jul,
  pages     = {91454E},
  series    = {Proc. SPIE},
  title     = {{SPEED: the segmented pupil experiment for exoplanet detection}},
  volume    = 9145,
  year      = 2014
}

@article{SPEED2015,
  adsurl  = {http://adsabs.harvard.edu/abs/2015Msngr.159...19M},
  author  = {{Martinez}, P. and {Preis}, O. and {Gouvret}, C. and {Dejongue}, J. and 
             {Daban}, J.-B. and {Spang}, A. and {Martinache}, F. and {Beaulieu}, M. and 
             {Janin-Potiron}, P. and {Abe}, L. and {Fantei-Cujolle}, Y. and 
             {Ottogalli}, S. and {Mattei}, D. and {Carbillet}, M.},
  journal = {The Messenger},
  month   = mar,
  pages   = {19-22},
  title   = {{The SPEED Project: SPEEDing up Research and Development towards High-contrast Imaging Instruments for the E-ELT}},
  volume  = 159,
  year    = 2015
}
\bibliographystyle{spiebib} % makes bibtex use spiebib.bst

\end{document}